\documentclass[useAMS,usenatbib]{mn2e}            
\usepackage{journals}
\usepackage[dvips]{graphicx,color}       
\usepackage[latin1]{inputenc}
\usepackage{graphics}    
\usepackage{amsfonts}    
\usepackage{amsmath}
\usepackage{multicol}     
\usepackage{layout}     
\usepackage{amssymb}
\usepackage[a4paper,colorlinks=true,pdfstartview=FitV,linkcolor=red,citecolor=blue,urlcolor=magenta]{hyperref}
 \title[Giant arc predictions]{Predicting the number of 
   giant arcs expected in the next generation wide-field 
   surveys from space} \author[Boldrin     et      al.      2012]
  {\parbox{\textwidth}{Michele   Boldrin$^{1}$,
    Carlo   Giocoli$^{1,2,3}$,  
    Massimo   Meneghetti$^{2,3}$, 
    Lauro  Moscardini$^{1,2,3}$\thanks{email: \href{mailto: miche.boldrin@gmail.com}{miche.boldrin@gmail.com}, \href{mailto: cgiocoli@oabo.inaf.it}{cgiocoli@oabo.inaf.it}, \href{mailto: massimo.meneghetti@oabo.inaf.it}{massimo.meneghetti@oabo.inaf.it}, \href{mailto: lauro.moscardini@unibo.it}{lauro.moscardini@unibo.it}}}   \\  \\  
  $^{1}$  Dipartimento di Astronomia, Universit\`a di Bologna,
  via  Ranzani  1,  40127,  Bologna,  Italy   \\ 
  $^{2}$ INAF  -  Osservatorio Astronomico di Bologna,  via Ranzani 1, 40127, Bologna,
  Italy \\  
  $^{3}$ INFN - Sezione di Bologna, viale Berti Pichat 6/2, 40127, Bologna, Italy}
\begin{document}

\date{}
\maketitle
\label{firstpage}
\pagerange{\pageref{firstpage}--\pageref{lastpage}} \pubyear{2012}
\begin{abstract}

  In  this  paper  we  estimate   the  number  of  gravitational  arcs
  detectable  in  a wide-field  survey  such  as  that which  will  be
  operated  by  the  Euclid  space mission,  assuming  a  $\Lambda$CDM
  cosmology.   We use  the  publicly available  code \textsc{moka}  to
  obtain  realistic  deflection  angle   maps  of  mock  gravitational
  lenses. The maps are processed by a ray-tracing code to estimate the
  strong lensing cross sections of  each lens.  Our procedure involves
  1) the  generation of  a light-cone which  is populated  with lenses
  drawn  from a  theoretical mass-function;  2) the  modeling of  each
  single lens using  a triaxial halo with  a NFW (Navarro-Frenk-White)
  density   profile  and   theoretical  concentration-mass   relation,
  including substructures,  3) the determination of  the lensing cross
  section as a  function of redshift for each lens  in the light-cone,
  4) the simulation of mock  observations to characterize the redshift
  distribution  of  sources that  will  be  detectable in  the  Euclid
  images.   We  focus   on  the  so-called  {\em   giant  arcs},  i.e.
  gravitational  arcs characterized  by  large length-to-width  ratios
  $(l/w\ge  5,7.5$ and  10).   We quantify  the  arc detectability  at
  different  significances   above  the   level  of   the  background.
  Performing  $128$ different  realizations  of a  15,000 sq.   degree
  survey,  we find  that the  number of  arcs detectable  at $1\sigma$
  above   the    local   background   will    be   $8912_{-73}^{+79}$,
  $2914_{-25}^{+38}$, and $1275_{-15}^{+22}$  for $l/w\ge5$, $7.5$ and
  $10$,   respectively.   The   expected  arc   numbers  decrease   to
  $2409_{-28}^{+24}$,  $790_{-12}^{+10}$,  and $346_{-6}^{+6}$  for  a
  detection limit at $3 \sigma$  above the background level.  From our
  analysis, we  find that most of  the lenses which contribute  to the
  lensing  optical depth  are located  at redshifts  $0.4<z_l<0.7$ and
  that the $50\%$  of the arcs are images of  sources at $z_s>3$. This
  is  the  first  step  towards   the  full  characterization  of  the
  population of strong  lenses that will be observed  by Euclid. Given
  these results, we conclude that  Euclid is a powerful instrument for
  strong lensing  related science,  which will  be useful  for several
  applications, ranging from  arc and Einstein ring  statistics to the
  measurement of the matter content in the cluster cores.

\end{abstract}

\section{Introduction}
It  is unanimously  recognized  that galaxy  clusters  are objects  of
enormous cosmological importance. The  reasons are manyfold. First, in
the framework of the  model of hierarchical structure formation, these
are the  latest structures  to form, being  the most  massive. Second,
they  sample the  exponential tail  of the  mass function,  thus their
abundance as a function of time is strongly sensitive to the values of
cosmological   parameters   \citep{press74,lacey93,sheth02}.    Third,
compared  to galaxies, their  internal structure  is less  affected by
baryons  due to their  small age,  which implies  that they  are ideal
laboratories  to  study  the  build  up of  the  cosmic  structure  as
predicted by the  theory of Cold Dark Matter  (CDM).  Several clusters
have been  captured in the  middle of a  merging phase leading  to the
formation  of a larger  structure. The  interplay between  baryons and
dark matter  during such  violent events highlights  the fundamentally
different behavior of the cluster mass components, providing wealth of
information on what  is the nature of dark  matter.  Alternatively, in
those systems which appear to be relaxed, other predictions of the CDM
theory can be tested, such as that the density profiles of dark-matter
halos should converge to a (nearly) universal law, as originally found
by \citet{NFW97}.

An increasing  number of observational campaigns  are targeting galaxy
clusters  with the  aim of  using  them as  cosmological probes.  They
explore  the whole wavelength  range where  clusters can  be detected,
from    the   X-ray    \citep{reflex,murray05},    to   the    optical
\citep{locuss,cccp},  to   the  radio  \citep{raccanelli12},   to  the
millimeter   \citep{swetz11,spt,planck}   wavelength  ranges.    These
observations complement  each other to  reach a comprehensive  view of
how visible and invisible  matter components are distributed.  Several
techniques  have been  developed to  this  goal. For  example, in  the
optical  domain, a direct  method to  investigate the  distribution of
matter  is via  the gravitational  lensing effect.  Due to  their huge
mass,  galaxy  clusters are  the  most  powerful gravitational  lenses
observable in  the sky. The  light emitted by distant  galaxies, which
happens  to  travel through  the  space-time  perturbed  by a  cluster
gravitational potential, is deflected,  causing a number of observable
effects, like distortions,  magnifications, and image multiplications.
Since galaxy clusters  extend over areas of many  arc minutes squared,
all lensing  regimes can be  observed around these  systems.  Galaxies
located at large angular distances  from the cluster center are weakly
but coherently sheared tangentially to the cluster. Sources at smaller
angular distances from  the cluster center can be  split into multiple
images and/or  appear as  very elongated arcs.  Both the weak  and the
strong lensing  regimes can be  inverted to derive information  on the
cluster mass distribution. In  particular, their combination allows to
derive precise  measurements of the  mass profiles from scales  of few
tens     of     kpc     to     the    virial     radius     \citep[see
e.g.][]{meneghetti10b,giocoli12b}.

Among those which  will investigate galaxy clusters  via their lensing
effects,  the survey  which will  be  operated by  the Euclid  mission
\citep{EUCRB}\footnote{\url{http://www.euclid-ec.org}}  is  likely  to
provide the highest quality data  and the widest sky coverage.  Euclid
has been recently selected by ESA  as medium-class space mission to be
launched in 2019.  It consists of  a $1.2$m space telescope which will
observe the sky both in the  optical and in the near-infrared domains.
In the optical, imaging will be taken  to the nominal depth of 24.5 AB
magnitudes in a wide $riz$  filter ($10\sigma$ extended sources).  The
telescope will be  sensitive to photons at  wavelengths between $0.55$
and $0.9  \mu$m. For the same  fields, Euclid will deliver  imaging in
the  $Y,J,H$  NIR bands  as  well  as  slitless spectra  covering  the
wavelength  range  $1.1-2.0 \mu$m.   The  main  survey which  will  be
operated by Euclid is a 15,000 sq. degree wide survey of the sky
at galactic latitudes  $|b|>30$ deg.  Euclid wants to  use two primary
probes  to investigate  the dark  components of  the Universe.   These
probes are weak-gravitational lensing  by the large-scale structure of
the       Universe,       Galaxy        Clustering       and       the
Baryonic-Acoustic-Oscillations.   However, within  the surveyed  area,
Euclid  is expected  to observe  several tens  of thousands  of galaxy
clusters.  With the telescope design being developed under the driving
requirement  of allowing  weak-lensing  measurements of  unprecedented
quality, lensing  will be the  ideal method to investigate  the matter
content of these structures.

Not  only Euclid  will be  a  powerful instrument  to investigate  the
weak-lensing  regime,  but it  will  also  deliver  high quality  data
suitable for  the strong lensing  analysis. Given the  high resolution
(0.1"/pixel)  and  sensitivity, Euclid  will  be  able  to detect  and
resolve  with  sufficient  accuracy  several strong  lensing  features
arising from highly magnified distant galaxies near the cluster cores,
and use them in combination with weak-lensing to extend the mapping of
the cluster content to the central regions.

Additionally,  strong lensing  will  be also  used  to constrain  both
cosmological parameters  and the  cluster properties following  a more
statistical approach.  It  has been shown by several  authors that the
number  of  strong lensing  features  on  the  whole sky  is  strongly
dependent on  1) the  geometrical properties of  the universe;  2) the
abundance of strong  lenses as a function of redshift;  and 3) several
cluster properties, such  as the average density  profile, the masses,
the  three-dimensional   shape,  the  concentration,  the   amount  of
substructures, the dynamical state, etc.  Strong lensing statistics is
therefore a  potentially powerful tool  to constrain cosmology  and to
trace the structure  formation \citep{wambsganss95,wambsganss98}. This
argument holds  in particular for  very rare features like  {\em giant
  arcs},  i.e.  gravitational  arcs characterized  by large  values of
their length-to-width ratio, $(l/w)\ge (l/w)_{\rm min}$.  The usage of
giant  arcs statistics  for  constraining the  cosmological model  was
proposed by \cite{B98}, who noted  that the expected abundance of such
events  differs by  orders-of-magnitude  between cosmological  models.
Intriguingly,  their simulations  leaded  to the  conclusion that  the
number  of  arcs  predicted  in  the  currently  favored  $\Lambda$CDM
cosmology  is about  one order-of-magnitude  smaller than  observed in
samples of X-ray selected galaxy clusters \citep{lefevre94,luppino99}.
Further investigations  on optically selected clusters  found that the
frequency of giant  arcs is particularly high at  redshifts $z \gtrsim
0.5$ \citep{zaritsky03,gonzalez03,horesh10}.

The mismatch between arc statistics  and other cosmological probes has
been tagged as {\em  the arc statistics problem} \citep{meneghetti03}.
A number of studies have been conducted with the aim of explaining its
origin \citep[e.g.][]{dalal2004,meneghetti2003,torri2004,puchwein2005,
  meneghetti2007,wambsganss2008,mead2010,horesh05,horesh11,li05,li06,
  wambsganss04,wambsganss05}, but  the controversy is not  yet solved.
Indeed,  it was  recently enforced  by several  other observations  of
strong lensing  clusters, which seem  to indicate that 1)  some galaxy
clusters have  very extended  Einstein rings  (i.e. critical  lines) {
  whose abundances} can hardly be  reproduced by cluster models in the
framework         of          a         $\Lambda$CDM         cosmology
\citep{tasitsiomi04,broadhurst2008b},  and  2) several  clusters,  for
which  high-quality strong-  and weak-lensing  data became  available,
have concentrations that are far  too high compared to the theoretical
expectations  \citep{broadhurst2008a,zitrin09}.  These  evidences push
in the same direction of the arc statistics problem, in the sense that
they both suggest that observed galaxy clusters are strong lenses that
are too effective compared to numerically simulated clusters.  Current
cluster surveys like CLASH (Cluster  Lensing And Supernova survey with
Hubble \cite{postman12}) are addressing these  issues and will help to
substantially understand the internal  structure and the peculiarities
of strong lensing clusters.

Existing  gravitational arc surveys  are based  on cluster  samples of
limited  size (few  tens of  galaxy  clusters). Euclid  will allow  to
dramatically increase  the number of  available strong lenses:  on the
basis of  simple extrapolations, it  is likely that thousands  of arcs
will be  detected in  the Euclid wide  survey \citep{EUCRB}.   In this
paper, we aim  at making robust estimates on the  number of giant arcs
which will be detectable in future Euclid observations. In particular,
we wish  to quantify the expected  number of such arcs  in a reference
WMAP-7  normalized  cosmological  framework  \citep{komatsu2011}  with
$\Omega_{0,m}=0.272$,    $\Omega_{\Lambda}=0.728$,    $h=0.704$    and
$\sigma_8=0.809$. The goal is to provide hints on what kind of cluster
strong-lensing related science will be possible with Euclid.

The paper is organized  as follows: in Sect.~\ref{method} we introduce
the relevant  lensing quantities and  we describe the methods  used to
compute  the number of  arcs.; in  Sect~\ref{results}, we  discuss the
results of the simulations,  focussing in particular on the dependence
of the  expected number of arcs  on the lens and  source redshifts and
finally, in Sect.~\ref{conclusions}, we draw our conclusions.

\section{Analysis}
\label{method}
\subsection{Synthetic halos}
Our theoretical  expectations are  based on the  analysis of a  set of
synthetic   halos  generated   with   the  public   code  {\tt   MOKA}
\cite{MOKA2011}. This code  was recently developed by us  with the aim
of  speeding up  strong lensing  calculations. It  is well  known that
accurate  estimates  of the  ability  of  clusters  to produce  strong
lensing  effects requires high  level of  details in  cluster modeling
\citep{meneghetti2000,meneghetti2003CD,torri2004,meneghetti2007,
  meneghetti2010}.  So far, the required complexity of the lens models
was ensured  only by numerical simulations, i.e.  by clusters obtained
from         N-body        and         hydrodynamic        simulations
\citep{meneghetti2003analytic,puchwein2005}.   {\tt  MOKA}  allows  to
create mock lenses using  a fast semi-analytic approach, through which
all the  cluster properties that  are relevant for strong  lensing are
incorporated   in   the   lenses.    This   is   achieved   by   using
simulation-calibrated analytical  relations to describe  the shape and
the content of  clusters.  A detailed description of  the code and its
implementation can be found in \cite{MOKA2011}. Very briefly:

\begin{itemize}
\item clusters are assumed to possess a triaxial dark matter halo. The
  axial-ratios  describing the  elongation  of these  halos are  drawn
  following  the prescriptions  of  \cite{jing2002}. To  each halo,  a
  random orientation is assigned;
\item dark matter  is distributed in the halos  such that the averaged
  azimuthal  density profile  resembles the  Navarro-Frenk-White (NFW)
  density  profile  \citep{NFW97}.   The  halo concentration  and  its
  dependence on mass and redshift  is modeled using the $c-M$ relation
  of \cite{zhao09}. A concentration  scatter is assumed, which is also
  based  on   the  analysis  of  numerically   simulated  dark  matter
  halos. These  typically show that  concentrations at fixed  mass are
  log-normally  distributed   with  a   rms    $\sim  0.25$,  almost
  independent of redshift;
\item dark matter substructures are added to the lens models according
  to the substructure mass function found by \cite{giocoli2010}. Their
  spatial  distribution is  modeled following  the  cumulative density
  distribution by  \cite{gao2004}.  Each substructure  is approximated
  with a truncated Singular-Isothermal-Sphere;
\item a central Brightest-Cluster-Galaxy  (BCG) is added at the center
  of  the  dark matter  halos.   The stellar  content  of  the BCG  is
  approximated by a \cite{hernquist1990} density profile. We take into
  account the influence of the presence  of the BCG on the dark matter
  distribution   near   the   halo   center  using   the   recipe   by
  \cite{blumenthal1986},  which analytically  describes  the adiabatic
  contraction. The influence of baryons  settled on the halo center on
  the surrounding dark matter distribution has been studied both using
  analytical  calculations and numerical  simulations, and  during the
  last years the problem has also been addressed from an observational
  point  of view  \citep{schulz10}. However  recently \cite{newman11},
  modeling  the triaxiality  of  Abell 383,  have  ruled out  baryonic
  physics  which serve  to steepen  the central  dark  matter profile.
  Nowadays  this phenomenology  is still  an open  debate both  from a
  theoretical  -- where  the dark  matter behavior  seems  to strongly
  depend on the gas physics and treatment of the simulations -- and an
  observational point  of view, and further investigations  are out of
  the purposes  of this paper. However  we want to stress  that in the
  light of  what has  been shown by  \cite{MOKA2011} in  comparing the
  strong lensing  cross sections of  triaxial haloes without  and with
  BCG plus  adiabatic contraction, we  expect to find a  difference of
  the order  of $5-10\%$ between  clusters with and  without adiabatic
  contraction.

\end{itemize}

The  lensing  properties  of  halos  generated with  {\tt  MOKA}  were
discussed by \cite{MOKA2011}. Since we want our modeled strong lensing
halos to  be as  similar as possible  to numerically  simulated galaxy
clusters,  we include  all features  that significantly  influence the
strong  lensing behavior  in our  computation.  We  stress  that these
features  are related  only to  the  dark matter  halo structural  and
geometrical nature,  and to the  BCG: the unique  significant baryonic
element for strong lensing analysis. In \cite{MOKA2011} is tested that
all  characteristics  listed  above   are  essential  for  an  optimal
reproducing    of   simulated    galaxy   clusters    strong   lensing
behavior. Finally, it  is also very important to  note that {\tt MOKA}
is very efficient and allows to quickly generate a lens model within a
few seconds of CPU time on  a powerful personal computer. Since we aim
at simulating  a (almost) full-sky  survey of strong  lensing clusters
and  at sampling  a large  number of  lines of  sight,  which requires
generating a large number of lenses, in this work we use {\tt MOKA} to
produce the  mass distributions  which are then  analyzed by  means of
ray-tracing methods.
 
\subsection{Ray-tracing simulations and  cross sections}
By  using {\tt  MOKA}, we  generate three-dimensional  cluster models,
which  we project  along  arbitrary lines-of-sight.   The  usage of  a
semi-analytic formalism  allows to quickly compute  for each projected
mass distribution its  deflection angle field on the  lens plane. This
is used to distort the images  of a large number of background sources
in  order to  compute the  lens cross  sections for  giant arcs.   The
methods  employed  to measure  the  cross  sections are  explained  in
details  elsewhere \citep[see  e.g.][]{meneghetti2000}. Here,  we only
summarized briefly the procedure.

We use the  lens deflection angle maps to trace  bundles of light rays
from the observer position back  to a source plane at redshift $z_{\rm
  s}=2$.   This  is populated  with  an  adaptive  grid of  elliptical
sources, whose spatial resolution increases toward the caustics of the
lens.   The caustics are  lines on  the source  plane along  which the
lensing magnification diverges. Therefore, those sources which will be
placed   near   the   caustics   will  be   characterized   by   large
magnifications. The  magnifications induced  by lensing can  either be
tangential  (near the {\em  tangential} caustic)  or radial  (near the
{\em  radial} caustic).  The  adaptive source  refinement artificially
increases the number of highly  magnified and distorted images. In the
following analysis,  a statistical weight, $w_i$, which  is related to
the spatial resolution  of the source grid at  the source position, is
assigned  to each  source. If  $a$ is  the area  of one  pixel  of the
highest resolution source  grid, then the area on  the source plane of
which the $i$-th  source is representative is given  by $A_i=aw_i$. By
collecting  the rays  hitting  each  source on  the  source plane,  we
produce  distorted images  of these  sources on  the lens  plane.  The
images are analyzed individually by measuring their lengths and widths
using the method outlined in \cite{meneghetti2000}.

We define the lensing cross section for giant arcs, $\sigma_{l/w}$, as
\begin{equation}
	\sigma_{l/w}=\sum A_i \;,
\end{equation}
where the  sum is extended  to all sources  that produce at  least one
image with $(l/w)\ge (l/w)_{\rm min}$.

\subsection{Redshift evolution of the cross sections}

The  cross section  is sensitive  to  several lens  properties and  it
depends on the  cosmological parameters and the redshifts  of the lens
and of the  sources. If we pack all the  relevant lens properties into
the  vector of  parameters $\vec{p}$  and the  cosmological parameters
into  the vector  $\vec{c}$, then  the  expected number  of arcs  with
$(l/w)\ge (l/w)_{\rm min}$  and surface  brightness larger than  $S$ that
the lens can produce  is given by \begin{equation} N_{l/w}(\vec p,\vec
  c,z_l,S)=\int_{z_l}^{\infty}\sigma_{l/w}(\vec     p,\vec    c,z_{\rm
    l},z_{\rm s})n(z_{\rm s},S)\mbox{d}z_{\rm s}\;,
 \label{Narcs1halo}
\end{equation}
where  $z_{\rm  l}$  and $z_{\rm  s}$  are  the  lens and  the  source
redshifts, respectively, and $n(z_{\rm s},S)$ is the number density of
sources with  surface brightness larger  than $S$ at  redshift $z_{\rm
  s}$.

As explained  above, we  measure the lens  cross sections for  a fixed
source redshift,  $z_{\rm s}=2$. The  previous formula shows  that the
cross sections need to be  measured at all redshifts above $z_{\rm l}$
in order to calculate the number  of arcs expected from a single lens.
In  principle, this  would imply  running ray-tracing  simulations for
many source planes, which is computationally very demanding, given the
number   of   lenses  we   are   using   in   this  work.    Following
\citet{meneghetti2010},  we  prefer  to  determine  a  \textit{scaling
  function} to  describe the redshift evolution of  the cross section.
To construct this scaling function we proceed as follows.

\begin{figure}
 \centering
 \includegraphics[width=\hsize]{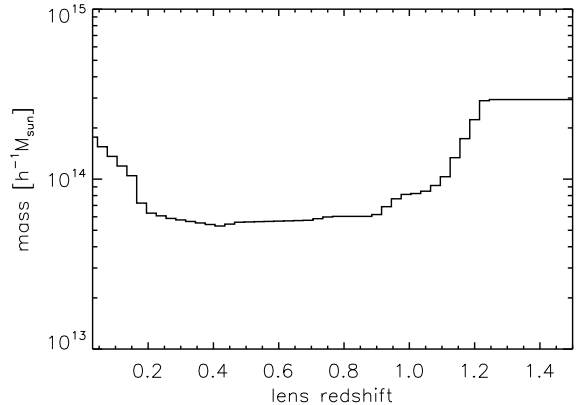}
 \caption{Minimal  mass for  producing  giant arcs  as  a function  of
   redshift,     as     derived     from    the     simulations     by
   \citet{meneghetti2010}.}
 \label{minmaxmasses}
\end{figure}

Although $\sigma_{l/w}$ depends on  a large number of lens properties,
$\vec{p}$,  we  can  identify   the  mass  as  the  primary  parameter
characterizing the  lens. Then, fixing the  cosmological framework, we
can write: \begin{equation} \sigma_{l/w}(M,z_{\rm l},z_{\rm s}) \equiv
  \langle    \sigma_{l/w}(\vec   p,\vec    c,z_{\rm    l},z_{\rm   s})
  \rangle_{\vec{\tilde{p}}}  \;, \end{equation}  where the  average is
taken  over  the remaining  lens  properties, $\vec{\tilde{p}}$  (i.e.
substructure content, concentration,  triaxiality and orientation). We
start  by producing  halos with  {\tt MOKA}  spanning three  orders of
magnitude in mass, in  the range $[10^{13}-10^{16}]$, distributed over
the  redshift  interval  $[0-1.5]$.    Halos  are  subdivided  in  100
logarithmically  equi-spaced  mass bins  and  50 linearly  equi-spaced
redshift bins. In  each ($M,z_{\rm l}$) cell, we  generate $100$ halos
with varying properties, $\vec{\tilde{p}}$, to be used for ray-tracing
simulations as  explained above.  Therefore,  the number of  lenses we
should  process is $100\times50\times100=500,000$,  which is  huge and
computationally  very  demanding.  The  numerical  study performed  by
\citet{meneghetti2010}  shows that  there  is a  minimal mass  $M_{\rm
  min}(z_{\rm l})$ at each redshift  below which halos are not capable
to produce giant arcs. To  reduce the computational time, we use their
results to  avoid the computation of  the cross section  of halos with
$M(z_{\rm  l})<M_{\rm  min}(z_{\rm  l})$,  for which  we  assume  that
$\sigma_{l/w}=0$. This allows us to  the reduce the number of halos to
be processed  using ray-tracing to  $\sim 340,000$.  The  minimal mass
adopted  in  our  study  is   shown  as  a  function  of  redshift  in
Fig.~\ref{minmaxmasses}.

\begin{figure}
 \centering
 \includegraphics[scale=0.49]{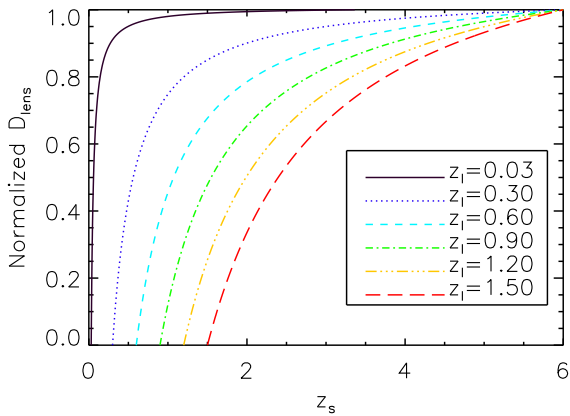}
 \caption{Normalized  lensing distance as  a function  of $z_s$  for 5
   different $z_l$ values, as shown in the label.\label{distlensing1}}
\end{figure}

We  measure $\sigma_{l/w}(M,z_{\rm l},z_{\rm  s}=2)$ by  averaging the
cross sections of  all halos in the ($M,z_{\rm  l}$) cell. This allows
us to obtain $\sigma_{l/w}(M,z_{\rm l},z_{\rm  s}=2)$ on a grid in the
($M,z{\rm l}$)  plane.  Then, we  use subsamples of 32  halos randomly
chosen in each  ($M,z_{\rm l}$) cell to repeat  the calculation of the
cross  sections for source  planes at  32 different  redshifts between
$z_{\rm  l}$ and $z_{\rm  s,max}=6$. These  source planes  are defined
such to take into account how rapidly the strong lensing efficiency is
expected  to to  grow  with  redshift. In  particular,  for each  lens
redshift $z_{\rm l}$, we use the {\em lensing distance} function
\begin{equation}
	D_{\rm lens}\equiv \frac{D_{\rm ls}D_{\rm l}}{D_{\rm {\bf s}}} \;,
\end{equation}  
where  $D_{\rm l}$  ,$D_{\rm  s}$  and $D_{\rm  ls}$  are the  angular
diameter distances  between the observer  and the lens  plane, between
the observer and the source plane, and between the lens and the source
planes, respectively.  We normalize  these functions such that $D_{\rm
  lens}(z_{\rm s}=6)=1$, and we  determine the redshifts of the source
planes  by  uniformly  sampling  the normalized  lensing  distance  at
intervals  $\Delta D_{\rm  lens}=1/32$. In  Fig.~\ref{distlensing1} we
show  the normalized  lensing distances  as a  function of  the source
redshift for several lens redshift. Our method to define the redshifts
of the source  planes ensures that many more  source planes are placed
in the redshift range where  the lensing distance grows rapidly, while
less planes are placed where the $D_{\rm lens}$ function becomes flat.

\begin{figure}
 \centering
 \includegraphics[width=\hsize]{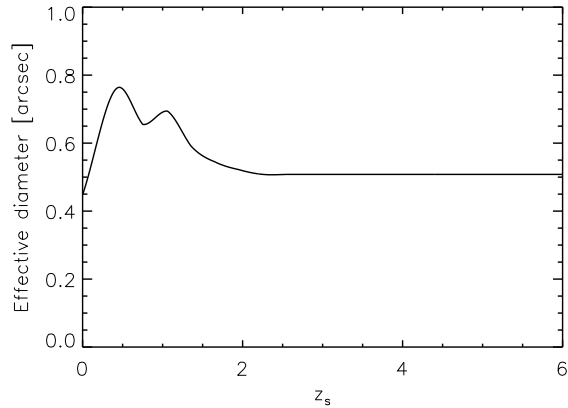}
 \caption{Apparent effective  diameter as  a function of  redshift, as
   found by \citet{gao2009}.}
 \label{sourcesize}
\end{figure}

A  critical   aspect  of  the  ray-tracing  simulations   and  of  the
measurement of the cross sections may  be given by the assumed size of
the  source galaxies,  which is  redshift  dependent.  \citet{gao2009}
studied how strongly  the lensing cross sections depend  on the source
sizes. They  found that this dependency  is very weak.  However, as it
does not delay the computation time, we include in our simulations the
redshift evolution of  the galaxy sizes, which is  modeled as follows.
\citet{gao2009}  used COSMOS  data \citep{scoville07}  to  measure the
redshift evolution of  the galaxy effective diameter up  to redshift 3
(see  their  Fig.  1).   The  median effective  diameter  measured  by
\citet{gao2009}   as   a   function    of   redshift   is   shown   in
Fig.~\ref{sourcesize}.  The  curve has been extended to  redshift 6 by
assuming no  evolution of  the galaxy sizes  above $z=3$. We  use this
function for setting the size of the sources as a function of redshift
in our ray-tracing simulations.

\begin{figure}
 \centering
 \includegraphics[scale=0.49]{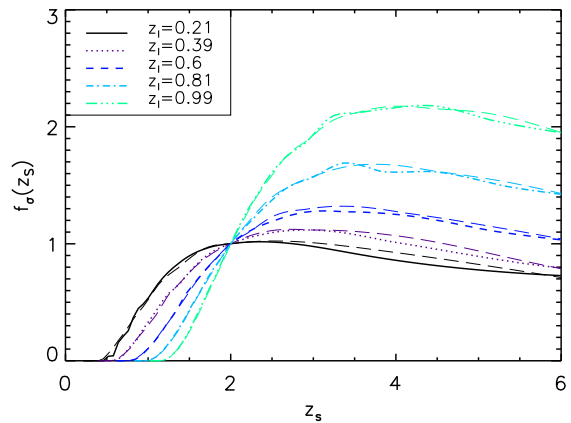}
\caption{Median scaling functions  derived from a sample  of 32 lenses
  with  $M\approx10^{15} h^{-1}M_{\odot}$  for five  $z_l$ values,  as
  shown in  the figure label.  The thick (long-dashed) lines  refer to
  functions computed without accounting for the source size dependence
  on redshift while thin lines are not.
   \label{distlensing2}}
\end{figure}

Having measured the cross sections for the different source planes, we
can     construct    the     scaling     functions    \begin{equation}
  f_{\sigma}(M,z_{\rm l},z_{\rm s})\equiv \dfrac{\sigma_{l/w}(M,z_{\rm
      l},z_{\rm s})}{\sigma_{l/w,0}(M,z_{\rm l},z_{\rm s}=2)}\;,
 \label{fsigma}
\end{equation}
where $\sigma_{l/w}(M,z_{\rm l},z_{\rm s})$  is estimated by averaging
over the 32 halos for each  source plane. Some examples of the scaling
functions  for halos  with  mass  $10^{15}h^{-1}M_{\odot}$ at  several
redshifts are  shown in the Fig.~\ref{distlensing2}.   By construction
all   scaling   functions   intercept    at   $z_{\rm   s}=2$,   where
$f_{\sigma}=1$. In Fig.~\ref{distlensing2}, the thin lines that almost
overlap  the  curves  represent  the scaling  functions  listed  above
computed  without  accounting  for   the  source  size  dependence  on
redshift.   As we  can see,  there is  no remarkable  difference among
curves, hence  we can  state that source  size dependence  on redshift
does not  significantly affect the  final number of arcs.   Anyway, as
already said,  adding this feature  does not change  the computational
time, so we decide to consider it in our implementation.

\begin{figure*}
 \includegraphics[scale=0.80]{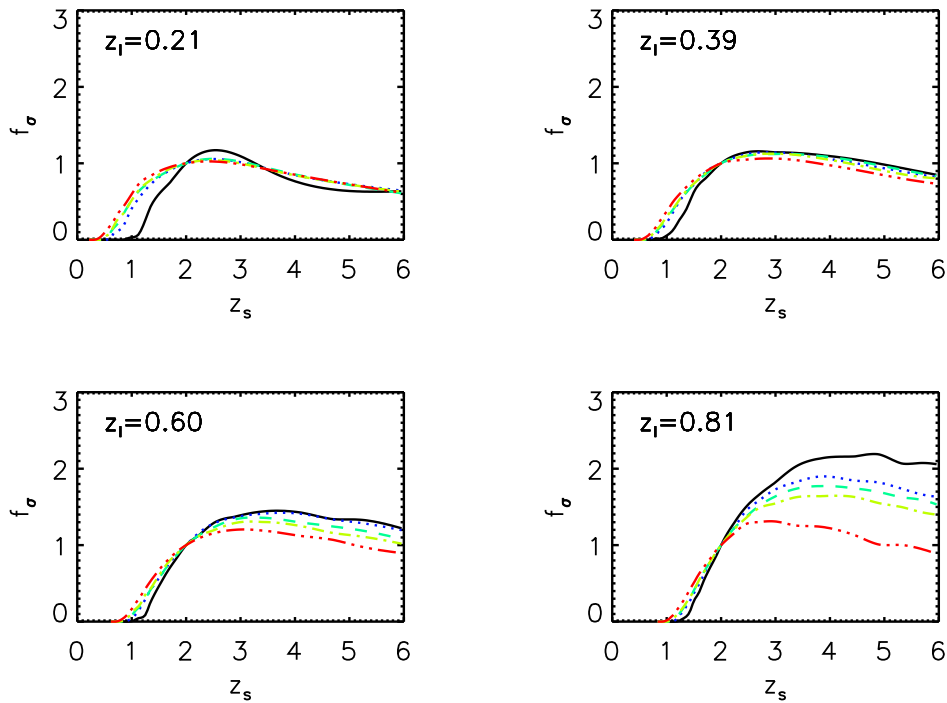}
 \caption{Scaling   functions  at   four  different   lens  redshifts.
   Starting from  the upper  left panel and  continuing to  the bottom
   right, the  results refer  to lenses at  $z_l=0.21, 0.39,  0.6$ and
   $0.81$, respectively. In each plot we show the curves corresponding
   to  five different  masses, namely  $2\times10^{14}h^{-1}M_{\odot}$
   (black solid line),  $4.5\times10^{14}h^{-1}M_{\odot}$ (blue dotted
   line),   $7.5\times10^{14}h^{-1}M_{\odot}$   (cyan  dashed   line),
   $10^{15}h^{-1}M_{\odot}$     (green     dot-dashed    line)     and
   $3.2\times10^{15}\,h^{-1}M_{\odot}$ (red double dot-dashed line).
 \label{scalingfunctions}}
\end{figure*}

Note that the scaling functions  depend not only on the lens redshift,
but    also    on    the    halo    mass.    This    is    clear    in
Fig~\ref{scalingfunctions}, which shows the scaling functions measured
at different redshifts  and for halos of different  mass. We see that,
at any  redshift, the scaling  functions for low-mass lenses  start to
rise at  larger $z_s$ compared to  lenses with higher  mass. They also
tend to reach  their maxima at significantly higher  redshift. This is
due to  the fact  that small lenses  are efficient at  producing giant
arcs  only  when  the  sources  are  distant.   Therefore,  it  is  of
fundamental importance to evaluate  the scaling functions in different
mass and redshift bins, as we do here.

By using  the scaling functions, we  can re-write Eq.~\ref{Narcs1halo}
as \begin{eqnarray}
  N_{l/w}(\vec p,\vec c,z_l,S) & = & \sigma_{l/w}(\vec p,\vec c,z_{\rm l},z_{\rm s}=2) \times \\
  &   &    \times   \int_{z_l}^{\infty}f_{\sigma}(M,z_{\rm   l},z_{\rm
    s})n(z_{\rm s},S)\mbox{d}z_{\rm s} \;,
 \label{1halo}
\end{eqnarray}
which allows  us to estimate the  number of arcs produced  by any lens
for  a given number  density of  sources just  by measuring  its cross
section at $z_{\rm l}=2$.

\subsection{Source number density}
The last  ingredient needed in  Eq.~\ref{1halo} to be able  to compute
the number  of giant arcs  expected from a  single lens is  the number
density  of sources  as a  function of  redshift and  limiting surface
brightness, $n(z_{\rm s},S)$.

For deriving the source redshift distribution function, we make use of
simulated    observations   with    the    {\tt   SkyLens}    software
\citep{meneghetti2008,meneghetti10b,bellagamba12,rasia12}.   This code
uses   a   set   of    real   galaxies   decomposed   into   shapelets
\citep{refregier03} to  model the  source morphologies on  a synthetic
sky. In particular, we use here  10,000 galaxies in the B, V,i,z bands
from  the Hubble-Ultra-Deep-Field  (HUDF)  archive \citep{beckwith06}.
Most galaxies have  spectral classifications and photometric redshifts
available \citep{coe06},  which are used  to generate a  population of
sources whose luminosity and  redshift distributions resemble those of
the HUDF. {\tt SkyLens} allows us to mimic observations with a variety
of telescopes, both from space and  from the ground. For this work, we
simulate wide-field observations with the optical camera which will be
onboard the  Euclid satellite.  For  setting up these  simulations, we
stick to the Euclid  description (throughput, PSF, telescope size, CCD
characteristics, etc.) contained in the Euclid Red-Book \citep{EUCRB}.
More details on Euclid simulations  carried out with the {\tt SkyLens}
software can be found in \cite{bellagamba12}.

We  simulate $400"\times  400"$  fields  to the  depth  which will  be
reached  by Euclid  ($m_{riz}\sim  24.5$), and  we  derive the  number
density and the  redshift distribution of all sources  detected in the
simulated  images. To  analyze the  images, we  use the  software {\tt
  SExtractor}  \citep{bertin96}, which  we  use also  to estimate  the
background rms. We derive source catalogs imposing different detection
thresholds, i.e $1$ and $3$ times the background rms.

\begin{figure}
 \centering
 \includegraphics[scale=0.50]{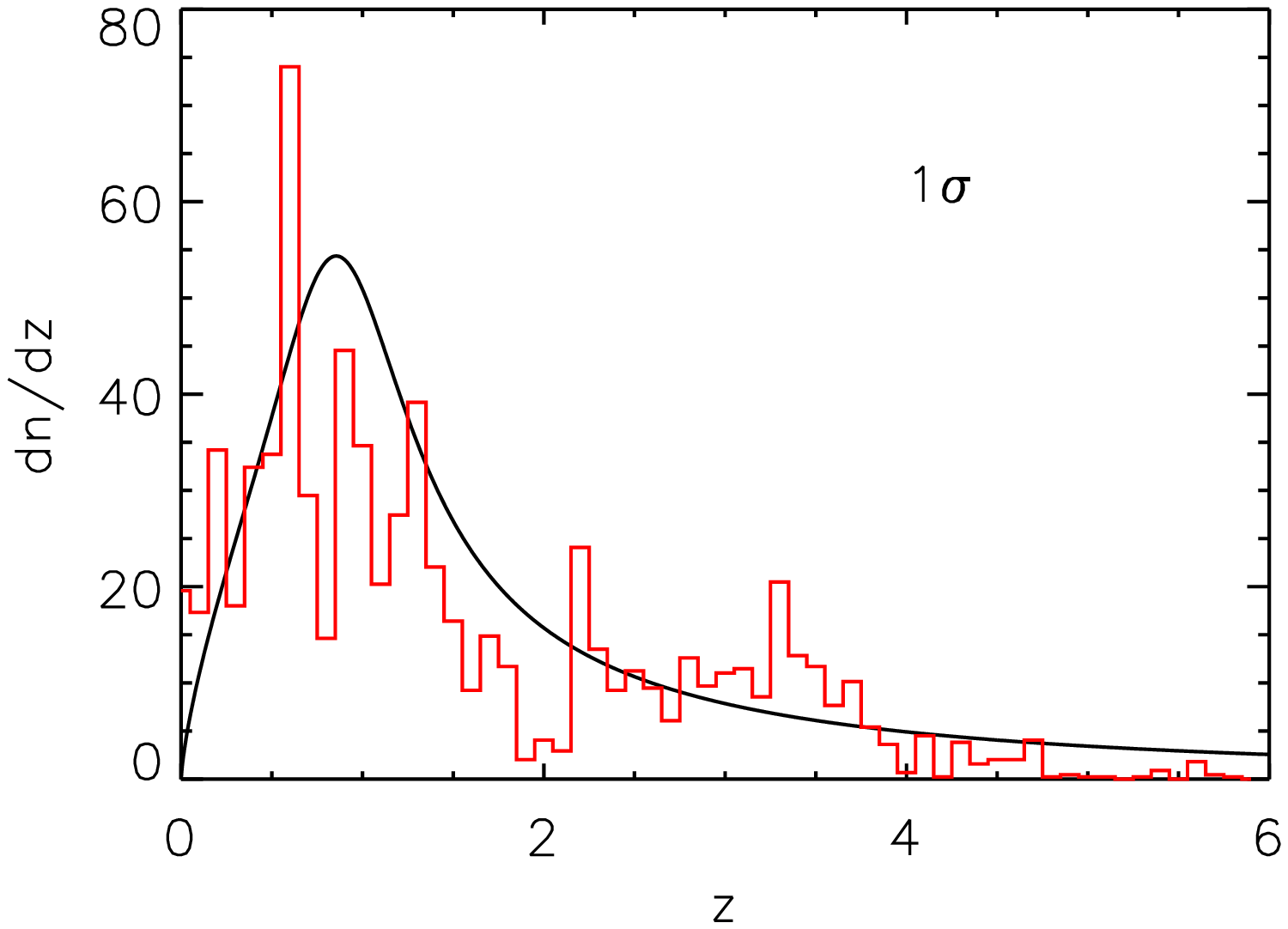}
 \includegraphics[scale=0.50]{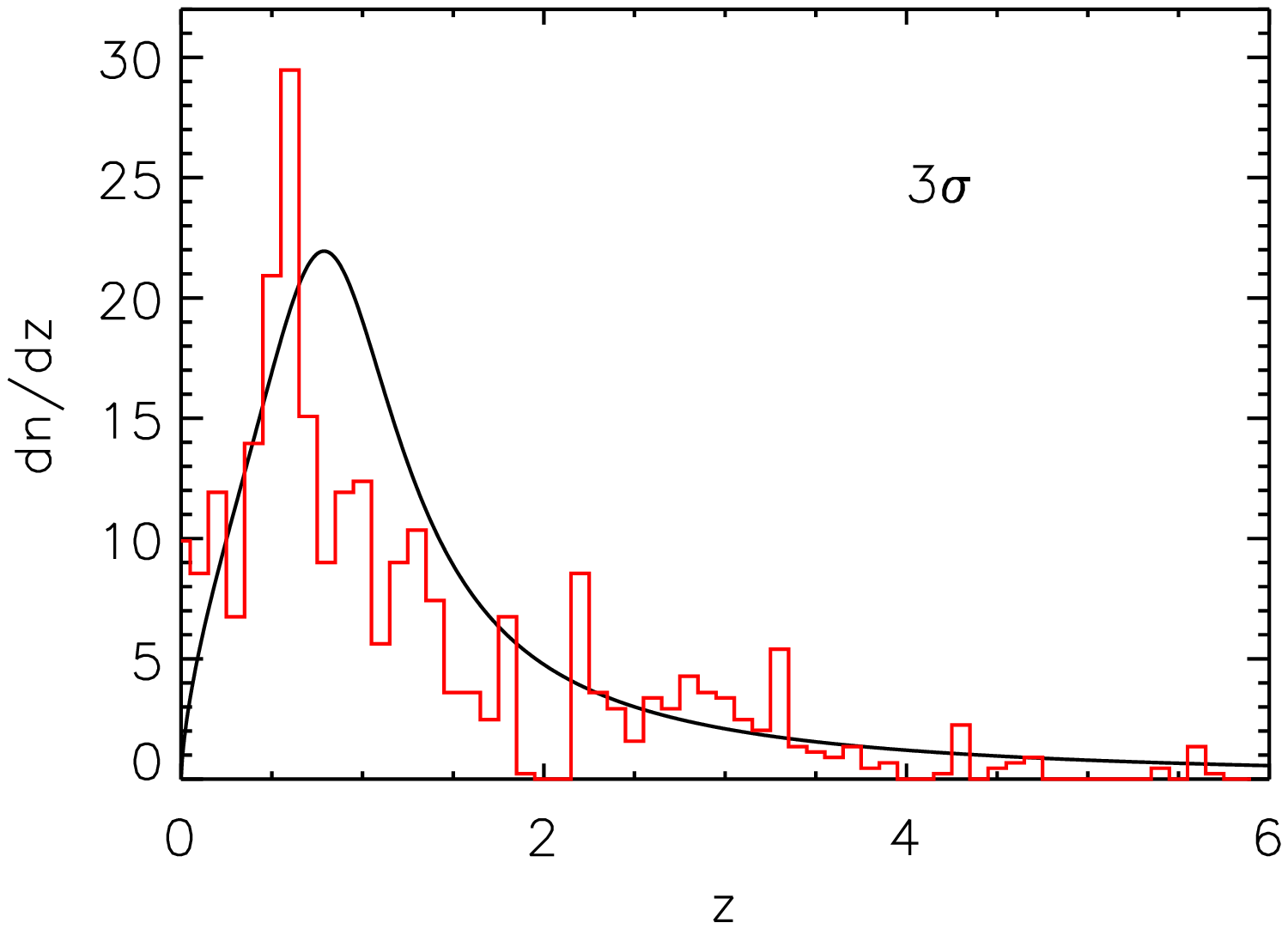}
 \caption{Source density  distribution as  a function of  redshift for
   galaxies  detected  at  1$\sigma$  and  3$\sigma$  above  the  mean
   background level. The red histograms show the distributions derived
   from the  analysis of the Euclid simulated  observations. The solid
   lines show the  best fit to the distributions  using the functional
   proposed by \citet{fu2008}. Numbers in  the $y$ axis are in unit of
   arcmin$^{-2}$.}
 \label{fits}
\end{figure}

The redshift distributions obtained for these two detection limits are
shown by the  histograms in Fig.~\ref{fits}, where we  plot the number
density of  detected sources as a  function of their  redshift. We fit
these  distributions with the  functional proposed  by \citet{fu2008},
which  has the  form  \begin{equation} n(z)=A\dfrac{z^a+z^{ab}}{z^b+c}
  \;,
 \label{functional}
\end{equation}
with
$$A=\left(\int_{0}^{+\infty}\dfrac{z^a+z^{ab}}{z^b+c}\mbox{d}z\right)^{-1}$$
and  $a,b,c$  free  parameters.   We  find  that  the  {\em  observed}
distributions are  fitted by  the functional with  bets-fit parameters
$(a,b,c)=(0.764,5.998,0.751)$  and  $(a,b,c)=(0.662,5.502,0.633)$  for
sources   $1\sigma$  and   $3\sigma$   above  the   mean  sky   level,
respectively.   These  best fits  are  shown  by  the solid  lines  in
Fig.~\ref{fits} from the same figures.

\subsection{Construction of the light-cones}
\label{buildingcones}
The procedure outlined above describes how we can calculate the number
of  arcs with  a given  $l/w$  ratio produced  by a  single lens.   By
investigating all lenses on our $(M,z_{\rm l})$ grid, we end up with a
list of $\sim 340,000$ cross  sections for sources at redshift $z_{\rm
  s}=2$, which we  can transform into cross sections  for other source
redshifts  using   the  previously  defined   scaling  functions.   In
particular, for each cell of the grid, we have $100$ cross sections of
halos with similar mass but different structural properties.

In  this  section, we  explain  how we  estimate  the  number of  arcs
expected  in a  given  area of  the  sky.  To  achieve  this goal,  we
obviously  need to  consider  all lenses  within  the light-cone  with
vertex  on  the observer,  which  subtends  the  surveyed area.   More
specifically,  aiming at  simulating  the wide  survey  which will  be
operated  by Euclid, we  construct light-cones  subtending an  area of
15,000 squared degree.   The depth of the light-cones  should be such
to contain all lenses capable  to produce giant arcs. According to the
simulations by  \citet{meneghetti2010}, we expect  no lenses producing
giant arcs from sources at  $z_{\rm s}=2$ above $z_{\rm l}\sim1.3$. To
be  more conservative, given  that our  simulations use  source planes
until redshift $z_{\rm s}=6$, we  extend the light-cones up to $z_{\rm
  l}=1.5$. It is worth mentioning,  however, that a giant arc has been
recently  discovered behind  the galaxy  cluster IDCS  J1426.5+3508 at
$z=1.75$ using  deep HST/ACS+WFC3 observations  \citep{gonzalez12}. On
the basis of  the arc color, the arc redshift  has been constrained to
be at  $z<6$, most likely  $z\sim4$.  The integrated magnitude  in the
F814W ACS filter is $24.29\pm0.31$,  thus close to the detection limit
of Euclid. As we will show later, in our simulations no giant arcs are
produced by lenses at $z_{\rm  l}>1.3$.  Thus, our results confirm the
peculiarity  of  this  arc   detection,  which  may  have  interesting
cosmological implications \citep{gonzalez12}.

Once defined the size of the light-cones, we populate them with lenses
with different  mass and redshift. To  do so, we divide  the cone into
$50$ redshift  slices, equi-spaced  in redshift with  $\Delta z=0.03$.
This  is  the  same  redshift  spacing  used  to  construct  the  grid
$(M,z_{\rm l})$ over which the cross sections were evaluated. Thus, we
define 50 lens  planes, with the first plane at  $z=0.03$ and the last
plane placed at redshift $1.5$.

\begin{figure}
 \centering
 \includegraphics[scale=0.50]{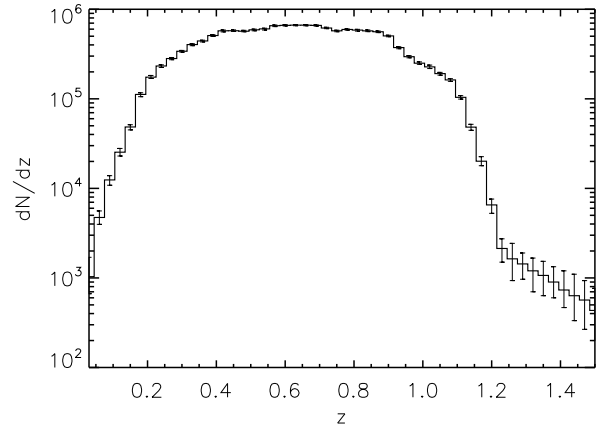}
 \caption{Median number of halos as  a function of redshift in the 128
   realizations  of the light  cone corresponding  to the  Euclid wide
   survey. The  error-bars indicate the minimum and  maximum number of
   halos  in  each  redshift   bin,  again  from  the  128  light-cone
   realizations.}
 \label{massdistr}
\end{figure}

We calculate the  number of the lenses with a given  mass to be placed
on  each  lens  plane by  using  the  Sheth  \& Tormen  mass  function
\citep{sheth99}.    Masses   are   drawn   again   in   the   interval
$[10^{13},10^{16}]\,h^{-1}M_{\odot}$.   To consider effects  of cosmic
variance, we  produce 128 realizations of the  light-cone. The average
number of halos generated in each redshift slice is shown by the solid
line  in  Fig.~\ref{massdistr},  where  the  error-bars  quantify  the
scatter between the different light-cone realizations.

To calculate the number of giant arcs expected to be detectable in the
surveyed area, for each halo of  mass $M$ and redshift $z_{\rm l}$, we
randomly select one  of the $100$ cross sections  in the corresponding
$(M,z_{\rm l})$ cell. Then, we assign to the halo the scaling function
previously  measured for  halos with  its  mass and  redshift. We  use
Eq.~\ref{1halo} to compute the number of arcs expected from each lens.
The total number of arcs expected in the survey is then calculated as
\begin{equation}
	N_{l/w}^{tot}=\sum_{i=1}^{N_{lens}}N_{l/w,i}\;,
\end{equation}  
where $N_{lens}$  is the  total number of  arcs in the  light-cone and
$N_{l/w,i}$ is the number of arcs produced by the $i$-th lens.

\begin{table*}
 \centering
 \begin{tabular}{l|l||c|c|c|c|c||}
	  &		 &$N_{med}$&I quartile&III quartile&$N_{min}$&$N_{max}$\\
\hline
\hline
   $l/w\ge5$&	$1\sigma$&8912	   &8839     &8991	 &8623	  &9308	     \\
	    &	$3\sigma$&2409	   &2381     &2433	 &2294	  &2482	     \\
\hline
   $l/w\ge7.5$&	$1\sigma$&2914	   &2889     &2952	 &2810	  &3100	     \\
	      &	$3\sigma$&790	   &779      &800	 &746	  &819	     \\
\hline
   $l/w\ge10$&	$1\sigma$&1275	   &1260     &1297	 &1216	  &1387	     \\
	     &	$3\sigma$&346	   &340      &352	 &323	  &362	     \\

 \end{tabular}
 \caption{$N_{med}$ is the median number of arcs with $l/w\ge5$,
   $7.5$, and $10$, computed from the results of 128 different 
   $15,000$ deg$^2$ mock light-cone realizations, from sources 
   $1\sigma$ and $3\sigma$ upon the mean sky level. In fourth and fifth 
   columns are the 25\% and 75\% percentiles, while in sixth and seventh 
   columns are the minimum and maximum values.}
 \label{totnumber}
\end{table*}

\section{Results}
\label{results}
In this  section, we show the  results of our  analysis, discussing in
particular the expected  number of arcs in the  Euclid wide survey and
the number of arcs as a function of the lens and source redshifts.

\subsection{The total number of arcs}
\label{totalnumber}
The total  number of arcs  expected in the  Euclid wide survey  on the
basis  of  our  simulations  is given  in  Tab.   \ref{totnumber}  for
different  minimal length-to-width  ratios ($l/w_{\rm  min}=5,7.5,10$)
and  for  two detection  thresholds,  namely  $1$  and $3$  times  the
background  rms  These  values   represent  the  threshold  above  the
background for  which a  group of connected  pixels are  identified by
\texttt{SExtractor} \citep{bertin96}.  We report  the median number of
arcs derived from the 128 realizations of the light-cones ($N_{med}$),
as well  as the quartiles of  the distributions.  To allow  for better
quantification of the  cosmic variance, we also report  the minima and
the maxima of the distributions.
 
If we consider the detections  above the background  rms, the median
numbers of arcs with $l/w\ge5$, $7.5$ and $10$ are $8912_{-73}^{+79}$,
$2914_{-25}^{+38}$ and $1275_{-15}^{+22}$ respectively. If we consider
the  detections  at  higher  significance ($3$  times  the  background
rms)     the    respective    numbers    are    $2409_{-28}^{+24}$,
$790_{-12}^{+10}$ and  $346\pm6$. The quoted errors  correspond to the
inter-quartile  ranges  of the  distributions.  We  notice that  those
values  are  dependent somehow  on  the  source redshift  distribution
adopted, which  is consistent with the simulations  performed with the
Euclid telescope equipment. A source redshift distribution with a pick
shifted $10\%$ below or above our fiducial one produces a total number
of arcs which is $20\%$ smaller or larger.

We would  like to  stress that  these arcs  will be  {\em potentially}
detectable in the future Euclid wide survey. At this stage, we are not
considering several  practical difficulties  which may  complicate the
recognition of  gravitational arcs in real  observations. For example,
arcs can be easily confused with edge-on spiral galaxies or with other
elongated structures  on the  CCDs. Additionally,  arcs form  in dense
regions of cluster galaxies. Since these are typically very bright and
extended, arcs are frequently hidden  behind them. Aiming at analyzing
huge datasets  such as the data  that will be delivered  by Euclid, it
will be particularly important to  develop softwares for the automatic
detection  of  gravitational  arcs.   Few  such  tools  exist  already
\citep{alard06,seidel07,cabanac07,more12}   and   have   been   tested
extensively. In  a work in  progress, we are currently  addressing the
task of  quantifying the degree  of contamination and  completeness of
the arc catalogs delivered from these arc finders through the analysis
of simulated images.

Nevertheless,  these results  indicate  that Euclid  will  be able  to
detect  an unprecedented  number of  strong lensing  features  such as
giant arcs and arclets. These will represent a treasury for any future
study focusing  not only on arc  statistics but also  aiming at using
these features to  construct and calibrate lens models  and to map the
mass distribution in galaxy clusters.

\subsection{Arc production as a function of the lens redshift}
\label{arcvszl}
It is  interesting to  study the redshift  distribution of  the lenses
producing giant arcs. This is important to assess which lenses will be
better  constrained  by  strong  lensing  data.  Moreover,  given  its
sensitivity to the dynamical evolution of clusters, it is important to
understand  up to  which redshift  gravitational arcs  can be  used to
trace cluster evolution.

\begin{figure*}
 \centering
 \includegraphics[scale=0.49]{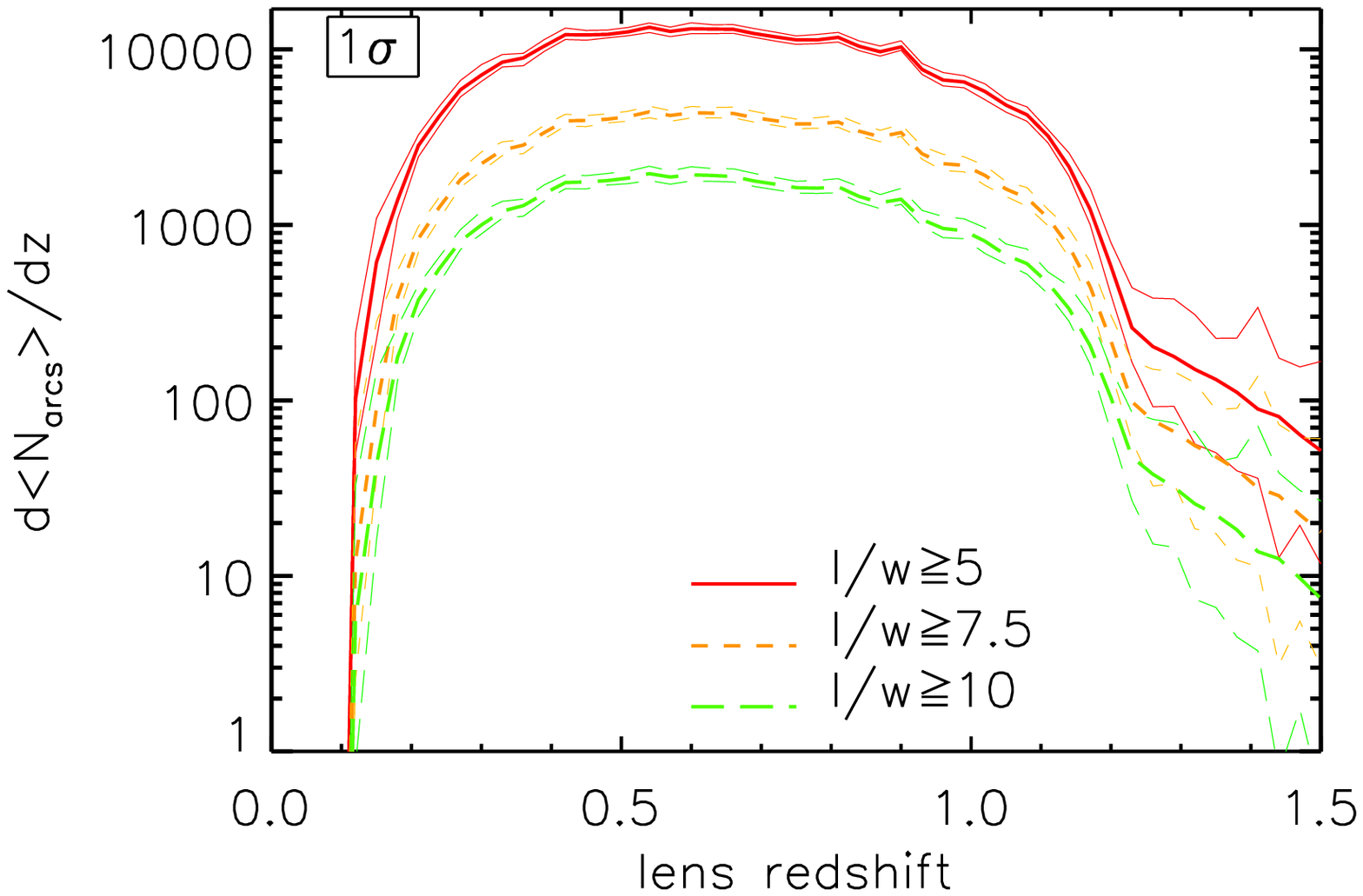}
 \includegraphics[scale=0.49]{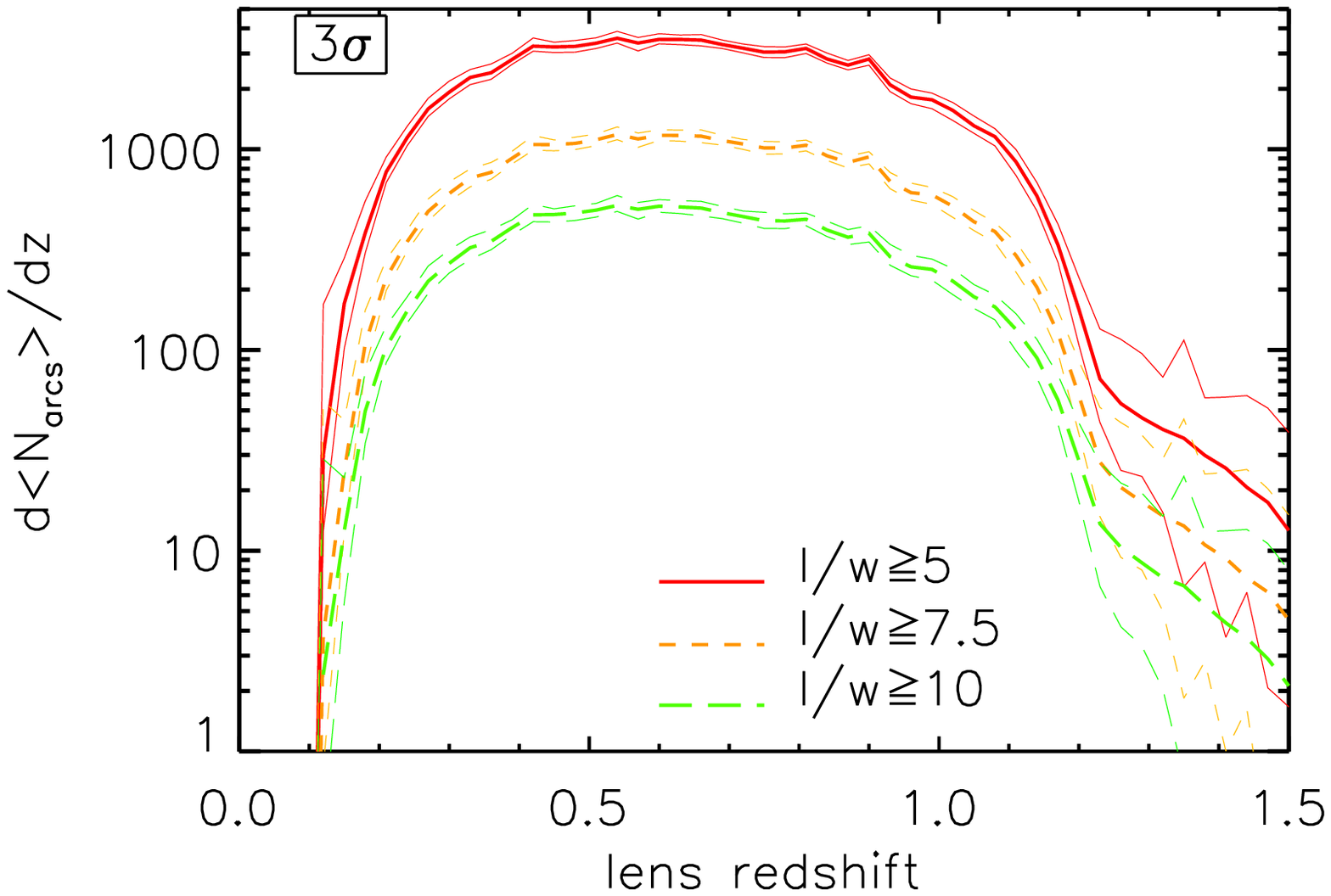}
 \caption{Number  of arcs  as a  function of  the lens  redshift.  The
   thick  (thin)  lines  are  the  median (quartiles)  among  the  128
   light-cone  realizations and they  refer to  arcs with  $l/w\geq 5$
   (solid red),  $7.5$ (dashed orange), and  $10$ (long-dashed green),
   respectively. The left and the  right panels refer to detections at
   the level of $1$ and $3$ times the background rms.}
 \label{narcszl}
\end{figure*}

In Fig. \ref{narcszl} we show the number of arcs produced by lenses at
different redshifts. We use solid  red, dashed orange and long-dashed
green lines to display the  results for arcs with $l/w\geq5$, 7.5, and
10, respectively. Shown are the medians of the 128 realizations of the
Euclid survey (thick lines) and the corresponding ranges among minimum
and maximum values  (thin lines). The left and  the right panels refer
to detections at the levels of $1$ and $3$ times the background  rms

\begin{figure}
 \centering
 \includegraphics[scale=0.43]{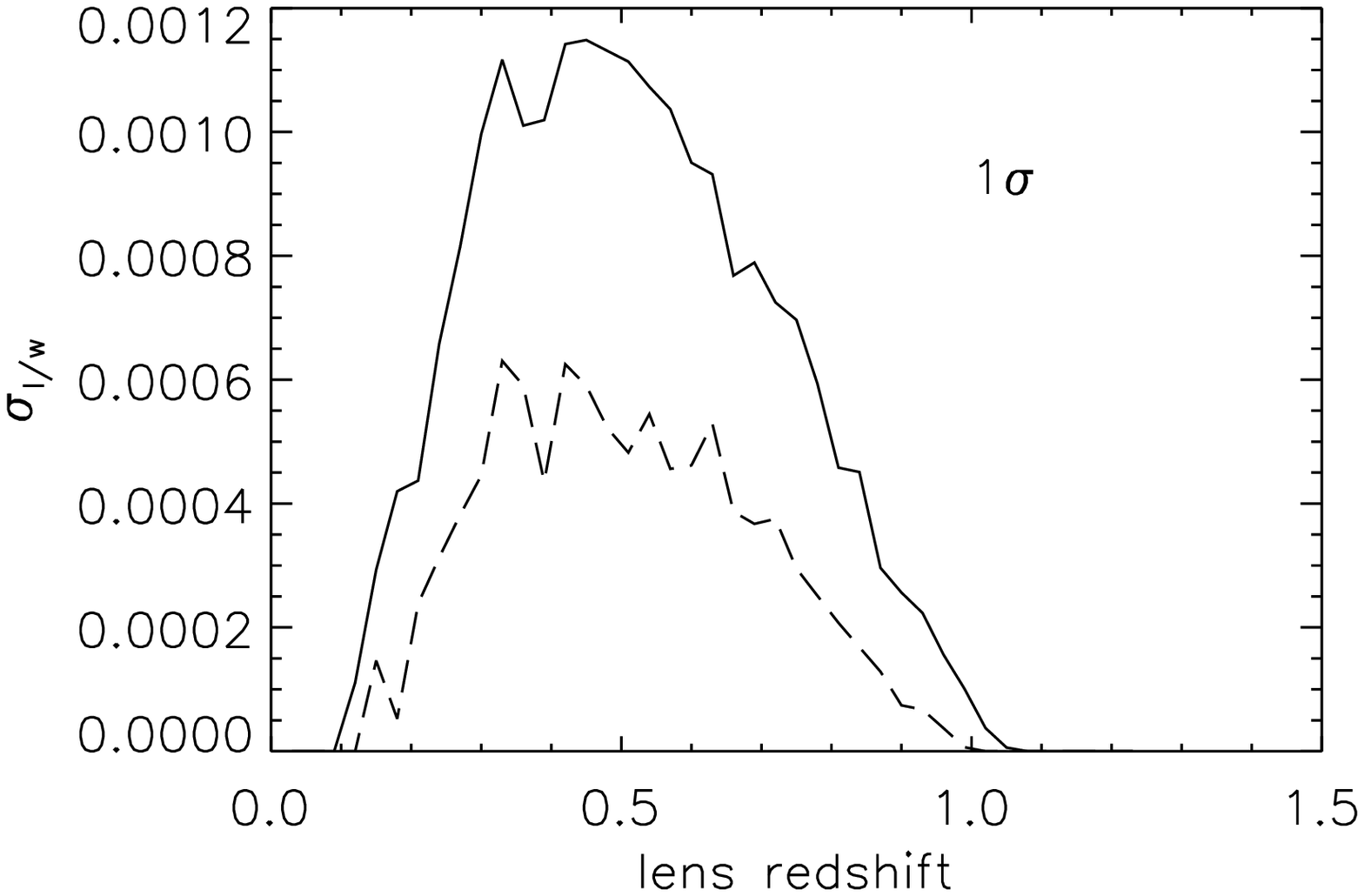}
 \includegraphics[scale=0.43]{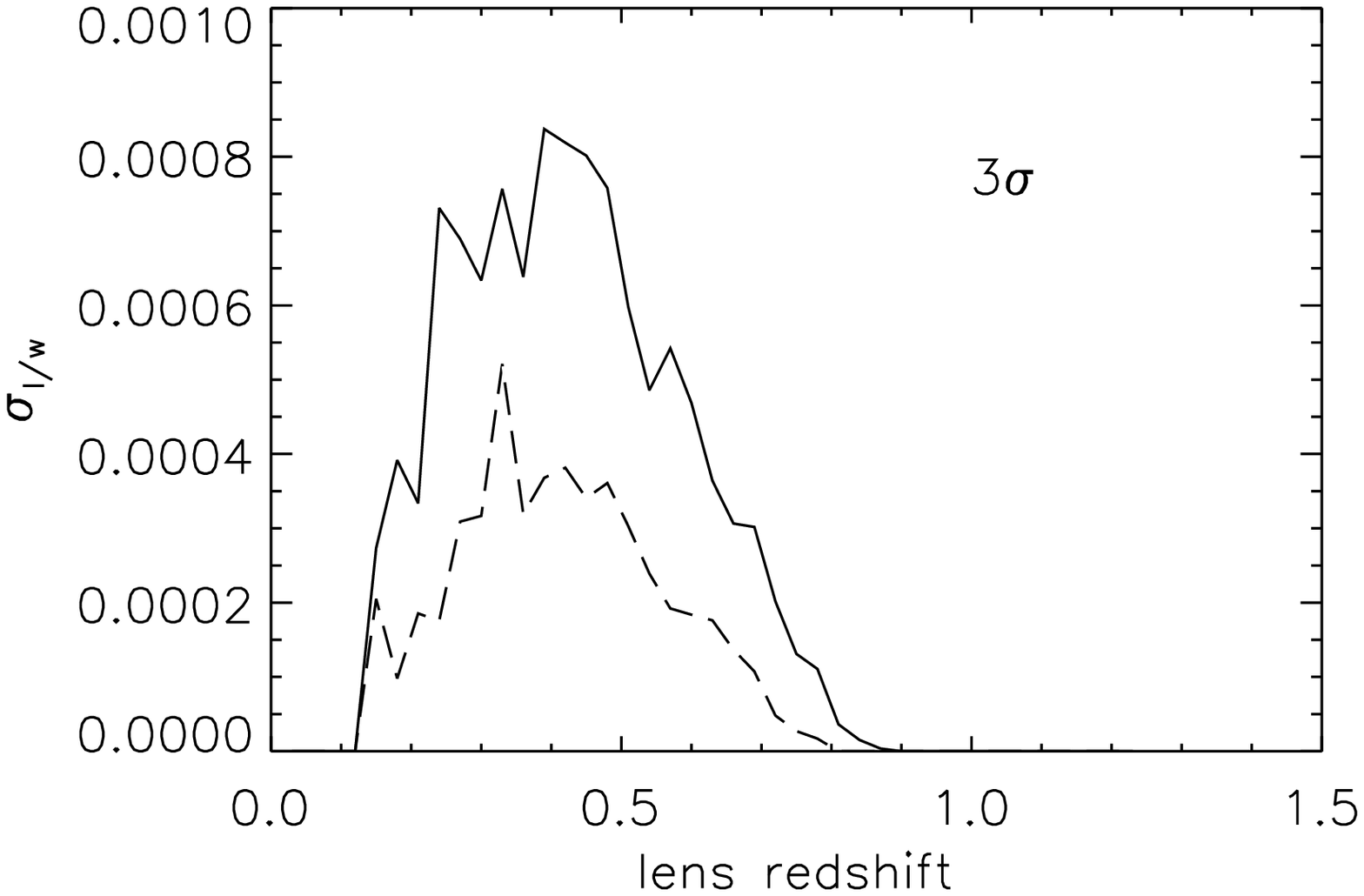}
 \caption{Lensing cross section as a function of the lens redshift for
   a halo  with mass $7\times 10^{14}h^{-1}M_{\odot}$.   The upper and
   the bottom panels  refer to detections at the level  of $1$ and $3$
   times the background  rms,  respectively. The solid and the dashed
   lines indicate the cross sections for arc with $l/w\ge7.5$ and with
   $l/w\ge10$,  respectively (cross sections  for arcs  with $l/w\ge5$
   have a similar behavior).}
 \label{crossseczl}
\end{figure}

We note that, independently of  the minimal $l/w$ ratio, the number of
arcs reaches its maximum at  redshift $\sim 0.6$.  It drops quickly to
zero at  redshifts $z\lesssim 0.2$ and $z\gtrsim  1.2$.  Such behavior
results  from  a combination  of  different  reasons.   First, at  low
redshift, the cosmic volume contained in the light-cone is small, thus
a relatively  small number of  lenses are present at  these redshifts.
This is  clear from Fig.~\ref{massdistr}, which shows  that the number
of halos  drops by almost two orders-of-magnitude  between $z=0.4$ and
$z=0.2$ and  by an  additional order-of-magnitude between  $z=0.2$ and
$z=0.1$.   Second, the lensing  cross section  of individual  halos is
small both  at low- and at  high-redshift, i.e.  when the  lens is too
close to the  observer or to the bulk of  sources. To illustrate this,
we show  in Fig.~\ref{crossseczl} the  lensing cross section  for arcs
with $l/w\geq7.5$  (solid lines) and  $l/w\geq10$ (dashed lines)  as a
function     of     redshift      for     a     halo     with     mass
$7\times10^{14}\;h^{-1}\;M_{\odot}$.  Given  the redshift distribution
of the sources expected in  the Euclid observations, the median source
redshift in  the case of arcs detectable  at the level of  $1$ and $3$
times  the background   rms  are  $z_s=z_{s,1\sigma}^{med}=1.24$ and
$z_s=z_{s,3\sigma}^{med}=1.03$, respectively. In  the upper and bottom
panels  of Fig.~\ref{crossseczl},  we  use these  source redshifts  to
calculate the cross sections. This explains why the curve in the upper
panel reaches its maximum at a slightly larger redshift than the curve
in the bottom panel.  Third,  as the redshift grows, increasingly less
massive  halos  are  expected,   which  implies  that  the  number  of
gravitational arcs  produced by  these lenses is  substantially lower.
Fourth,  although  high-redshift   sources  can  be  more  efficiently
distorted, their  surface brightness is  dimmed, and their  images are
more difficult to detect.

As we  can see from Fig.~\ref{crossseczl}, the  lensing cross sections
of each individual halo exhibit several local maxima at different lens
redshifts.  We  remind that  {\tt  MOKA}  produces  mock lenses  which
include substructures whose mass  and positions are drawn from recipes
calibrated  on numerical  simulations.   In particular,  halos may  be
produced with mass configurations resembling a merging phase. In fact,
the bumps  in Fig.~\ref{crossseczl}  correspond to such  events, which
are   known to boost  the lensing  cross section  and the  production of
arcs, \citep{torri2004} significantly. The same events are responsible
for the irregular behavior of the curves in Fig.~\ref{narcszl}.

\begin{figure*}
 \centering
 \includegraphics[scale=0.49]{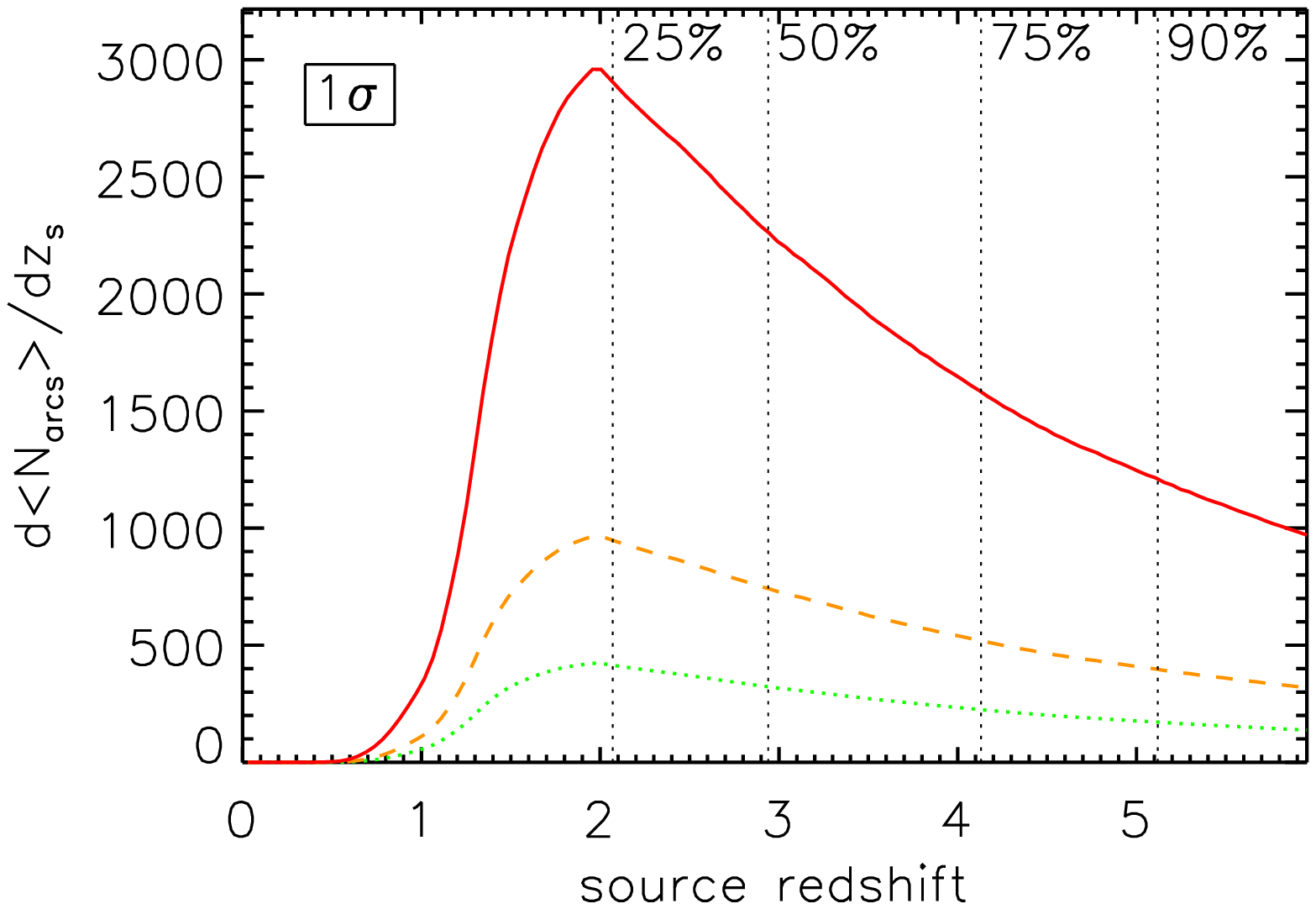}
 \includegraphics[scale=0.49]{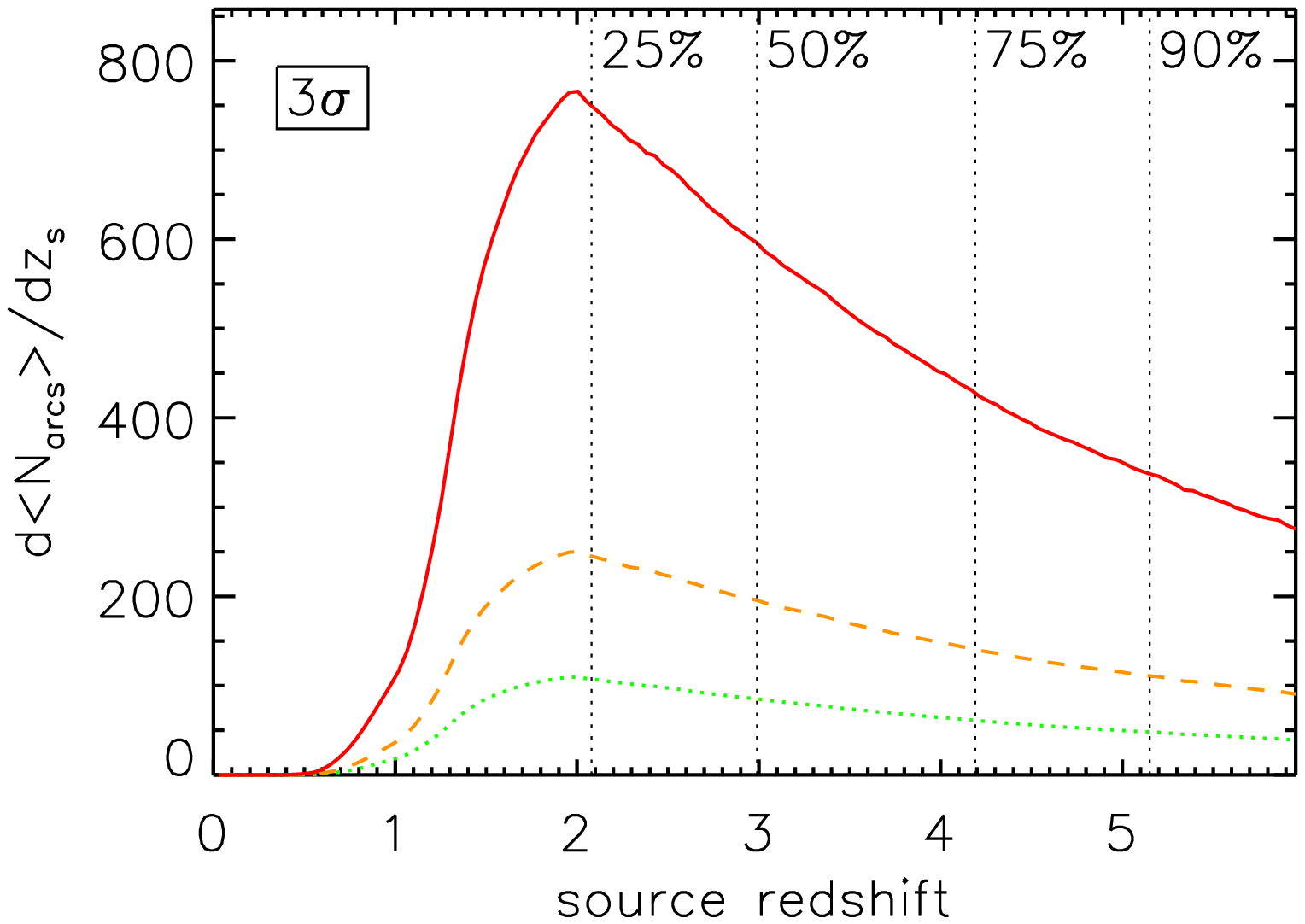}
 \caption{Number of  arcs as  a function of  $z_s$.  The left  and the
   right panels refer to detections at  the level of $1$ and $3$ times
   the background rms  Results are  shown for arcs with $l/w\geq 5$
   (solid  line), $7.5$ (dashed  line), and  $10$ (dotted  line).  The
   fractional numbers of arcs originated from sources at $z\leq z_{\rm
     s}$ are given by the  vertical dotted lines: they are independent
   of the value of $l/w$.}
 \label{narcszs}
\end{figure*}

\subsection{Arc production as function of the source redshift}
\label{arcvszs}

Now we discuss how the number of arcs expected from a Euclid-like wide
survey varies as a function of the source redshift.  This is useful to
understand what will be the  the typical redshift range of the sources
strongly  lensed  by   galaxy  clusters,  which  provides  interesting
information  on   how  accurately  the  lens  mass   profiles  may  be
constrained.

Our results  are summarized in  Fig.~\ref{narcszs}, where we  show the
number of arcs produced by all  lenses in the light-cone as a function
of the source  redshift. The solid, dashed, and  dotted lines refer to
arcs with $l/w\geq5$, $7.5$, and $10$, respectively. As usual, we show
the results  for detections  at the  levels of $1$  and $3$  times the
background rms (left and right  panels). The results shown here are
the medians over the  128 light-cone realizations. The vertical dotted
lines indicate the cumulative  fractional number of arcs originated by
sources  at  redshifts $\leq  z_{\rm  s}$.   All  curves, are  similar
independently of the value $l/w$,  have a maximum at $z_{\rm s}\sim2$,
indicating  that the  maximal efficiency  for  producing gravitational
arcs  is reached  at  this  redshift.  The  number  of arcs  decreases
quickly at lower redshifts. Only  $\sim25\%$ of the arcs is originated
by sources  at redshift $z_{\rm s}  \lesssim 2$.  On  the contrary the
vast majority  of arcs are  originated by sources at  higher redshift:
$\sim   50\%$  of   them  are   originated  by   sources   at  $z_{\rm
  s}\gtrsim3$. The  figure shows  that the arc  number is  expected to
decline gently as  a function of the source  redshift, with $25\%$ and
$10\%$ of them corresponding to sources at redshifts $\gtrsim 4.2$ and
$\gtrsim 5.2$, respectively.

Therefore, our results reveal that Euclid, due to its high efficiency,
will  be an  useful gravitational  telescope and  it will  improve the
study of distant galaxies \citep{zitrin12}.

\section{Conclusions}
\label{conclusions}
In this  work we outlined  a method to  calculate the number  of giant
arcs expected in  a wide survey. We particularized  our calculation to
the case  of the 15,000 sq.   degrees survey which will  be operated by
the Euclid space mission starting in 2019.  Our method is based on the
publicly available  code \texttt{MOKA} \citep{MOKA2011},  which allows
to  create  realistic  mock  galaxy  cluster  deflection  angle  maps,
including  all  features important  for  strong gravitational  lensing
features in a very short computational time. In particular it includes
triaxiality,  substructures, asymmetries, a  CD galaxy,  and adiabatic
contraction due to  baryons in cluster center.  We  recall that, since
it   uses   recipes  directly   calibrated   on  N-body   simulations,
\texttt{MOKA} produces models that are fully consistent with numerical
simulations \citep[see e.g.][]{MOKA2011}.

For our analysis,  we created a catalog of $\sim$  340,000 mock galaxy
clusters, spanning three  orders of magnitude in  mass and distributed
over the redshift interval $0 <  z < 1.5$, assuming a reference WMAP-7
normalized cosmology.  Using ray-tracing  techniques, we  measured the
strong lensing cross section for the  production of giant arcs of each
lens.  We also  estimated the evolution of the  lensing cross sections
as a function of the source redshift. We used the resulting catalog of
lensing cross  sections to generate  128 realizations of  the $15,000$
sq. degrees  survey, distributing mock lenses  in light-cones spanning
the  redshift  range $[0,1.5]$.   We  used  the \texttt{SkyLens}  code
\citep{meneghetti2008} to produce  realistic simulated observations by
Euclid, which  we used to  determine the redshift distribution  of the
galaxies which are likely to be lensed by foreground clusters.

With  so many  realizations of  the  Euclid survey,  we estimated  the
number  of  arcs,  and  its statistical  uncertainty,  which  will  be
detectable  in  future  Euclid  observations  at  different  detection
limits.  We discussed what is the typical redshift of lenses producing
arcs as well as what is the  redshift of sources that are lensed so as
to produce arcs detectable by Euclid. Our results can be summarized as
follows:
\begin{itemize}  

\item  Among the  different  realizations of  the  Euclid survey,  the
  median numbers  of arcs detectable  at the level of  $1\sigma$ above
  background  level  are  $8912_{-73}^{+79}$, $2914_{-25}^{+38}$,  and
  $1275_{-15}^{+22}$  for   arcs  with  $l/w\ge5$,   $7.5$  and  $10$,
  respectively;   such   numbers   decrease   to   $2409_{-28}^{+24}$,
  $790_{-12}^{+10}$ and $346\pm6$ for  arcs detectable at the level of
  three times  the local background rms  The  quoted errors reflect
  the first and the third quartiles of the arc number distribution.

\item Most  of the  arcs are  produced by lenses  at redshifts  in the
  range $[0.4,0.7]$. This  is due to the large  abundance of efficient
  lenses in this redshift range.

\item  We found  that  $50\%$ of  the  total number  of  arcs will  be
  produced by galaxies at redshifts $z\gtrsim 3$. Additionally, $25\%$
  and $10\%$ of  the arcs will be produced by  sources at $z>4.15$ and
  at $z>5.1$, respectively.  Thus, lensed sources in the future Euclid
  observations  will  be detected  up  to  very  high redshift.  Their
  redshift distribution  will peak at  $z\approx2$, almost independent
  of the length-to-width ratio.
\end{itemize}

Given these results, we conclude  that Euclid is a powerful instrument
for strong lensing  related science, which will be  useful for several
applications,  ranging from arc  and Einstein  ring statistics  to the
measurement of the matter content in the cluster cores.

Several works  in the  literature tried to  estimate how  the expected
number  of  arcs  reacts   to  changing  the  values  of  cosmological
parameters.   For example, \cite{meneghetti2003}  find that  in models
with  dynamical dark  energy we  should expect  of order  of $20-25\%$
larger  optical  depths for  strong  lensing  than  in a  $\Lambda$CDM
cosmology. \cite{2011MNRAS.415.1913D} show that variations of the same
order  of the  expected number  of arcs  are expected  in  models with
non-Gaussian initial  conditions ($-5\%$  to $+45\%$ for  f$_{\rm NL}$
varying between -10 and +74. \cite{fedeli2008} find that the number of
arcs  changes  by  one   order  of  magnitude  if  the  power-spectrum
normalization  $\sigma_8$ is  changed  from 0.7  to  0.9. Our  results
indicate that, if we consider  the statistical errors on the number of
giant  arcs alone,  Euclid will  be potentially  able  to discriminate
between  different  models of  dark  energy,  initial conditions,  and
cosmological  parameters. With  the procedure  we have  constructed in
this work, we can  easily compare different cosmological frameworks in
a reasonable time  and test whether future instruments  such as Euclid
will  be able  to use  arc  statistics as  an additional  cosmological
probe.

\section*{AKNOWLEDGEMENTS}
We thank  Thomas Kitching and  Henk Hoekstra for useful  comments.  We
are grateful also to  Matthias Bartelmann for interesting discussions.
Many thanks  also the the  anonymous referee for helpful  comments and
suggestions.   We acknowledge  financial contributions  from contracts
I/009/10/0, EUCLID-IC phase A/B1, PRIN-INAF 2009, ASI-INAF I/023/05/0,
ASI-INAF I/088/06/0, ASI I/016/07/0 COFIS, ASI Euclid-DUNE I/064/08/0,
ASI-Uni  Bologna-Astronomy Department  Euclid-NIS I/039/10/0  and PRIN
MIUR ``dark  energy and cosmology  with large galaxy  surveys''.  CG's
research has  been partially supported  by the project  GLENCO, funded
under  the  Seventh Framework  Programme,  Ideas,  Grant Agreement  n.
259349. The simulations of this project have been run during the Class
C Project-HP10CGR27W (\texttt{MOKAlen1}).

\bibliographystyle{mn2e}
\bibliography{biblioteca}
\label{lastpage}
\end{document}